# Doping of Ga in antiferromagnetic semiconductor α-Cr$_2$O$_3$oxide and its effects on modified magnetic and electronic properties

R.N. Bhowmik[1*], K.Venkata Siva[1], R.Ranganathan[2], and Chandan Mazumdar[2]

[1]Department of Physics, Pondicherry University, R. V. Nagar, Kalapet, Pondicherry-605014, India.

[2]Condensed Matter Physics division, Saha Institute of Nuclear Physics, 1/AF Bidhannagar, Kolkata-700064, India

[*]Corresponding author: Tel.: +91-9944064547; Fax: +91-413-2655734

E-mail: rnbhowmik.phy@pondiuni.edu.in

**Abstract**

The samples of Ga doped Cr$_2$O$_3$ oxide have been prepared using chemical co-precipitation route. X-ray diffraction pattern and Raman spectra have confirmed rhombohedral crystal structure with space group R$\overline{3}$C. Magnetic measurement has indicated the dilution of antiferromagnetic (AFM) spin order in Ga doped α-Cr$_2$O$_3$ system oxide, where the AFM transition temperature of bulk α-Cr$_2$O$_3$ oxide at about 320 K has been suppressed and ferrimagnetic behavior is observed from the analysis of the temperature dependence of magnetization data below 350 K. Apart from Ga doping effect, the spin freezing (50 K-70 K) and superparamagnetic behavior of the surface spins at lower temperatures, typically below 50 K, have been exhibited due to nano-sized grains of the samples. All the samples showed non-linear current-voltage (I-V) characteristics. However, I-V characteristics of the Ga doped samples are remarkably different from α-Cr$_2$O$_3$ sample. The I-V curves of Ga doped samples have exhibited many unique electronic properties, e.g., bi-stable (low resistance- LR and high resistance-HR) electronic states and negative differential resistance (NDR). Optical absorption spectra revealed three electronic transitions in the samples associated with band gap energy at about 2.67-2.81 eV, 1.91-2.11 eV, 1.28-1.35 eV, respectively.

## 1. Introduction

Recently, antiferromagnetic (AFM) systems (NiO, CoO, $Fe_2O_3$, $Cr_2O_3$) in nanoparticle (NP) form have received a significant amount of research interest for fundamental understanding of new magnetic phenomena, such as surface superparamagnetism, size-induced ferromagnetism, quantum confinement effect [1-7]. The enhanced ferromagnetism and magnetic exchange bias effect has provided the experimental evidence of modified spin order (core-shell structure) and magnetic anisotropy in AFMNP [2-4]. From application point of view, increase of ferromagnetic (FM) component in AFMNP is important for the development of magnetic semiconductor with reasonably large ferromagnetic moment, tunable band gap in the UV-Vis range of light and electrical field controlled electronic properties at room temperature [8-12].

Among the traditional AFM oxides, $\alpha$-$Cr_2O_3$ is a green colored compound that stabilizes in corundum (rhombohedral) structure with optical band gap $\sim 3$ eV [13] and used in the field of visible light driven photo catalysis, hydrogen storage, photonic and electronic devices and drug delivery [13-15]. Recently, some efforts have been made to develop magnetic semiconductors by doping magnetic Cr atoms into the lattices of wide band gap semiconductors ($Al_2O_3$, $Ga_2O_3$, ZnO) which can provide suitable materials for applications in spintronics and optoelectronics devices. For example, Cr doped $Al_2O_3$ have been used as thermo-graphic phosphor for the surface temperature measurement [16], Cr doped $Ga_2O_3$ have been used as a broadband laser material [17]. Cr doped ZnO has found potential applications in spintronics and optoelectronics [18]. Among the $Ga_2O_3$ oxides, $\beta$-form is a typical wide band gap semiconductor ($\sim 5$ eV). It has been used in gas sensors, field-effect transistors (FETs), and photo detectors [17]. The alloyed compound of corundum-structured α-M$_2$O$_3$ (M = Fe, Al, Cr) and Ga$_2$O$_3$ have shown unique magnetic and electronic properties useful for developing functional engineering materials [19]. The doping of non-magnetic Ga (3d$^{10}$ outer shell ionic configuration) in α-Fe$_2$O$_3$ (rhombohedral) structure by alloying of α-Fe$_2$O$_3$ and β-Ga$_2$O$_3$ has modified optical band gap, ferromagnetic, electronic and magneto-electronic properties [9, 11-12].

Considering a bright future of the metal doped corundum structured compounds, such as α-Fe$_2$O$_3$ and α-Cr$_2$O$_3$, for the development of new magnetic semiconductor, our objective in this work is to prepare the Ga doped α-Cr$_2$O$_3$ compound, and study the modification in structure, magnetic and electronic properties. We used current–voltage characteristics to highlight the non-linear electronic properties of the Ga doped Cr$_2$O$_3$ samples and future prospects of applying the material in electro-magnetic devices.

## 2. Experimental

The Ga doped Cr$_2$O$_3$oxide has been prepared by chemical reaction of the required amounts of Chromium Chloride (CrCl$_3$, purity ≥ 99.999%) and Gallium Chloride (GaCl$_3$, purity ≥ 99.999%) solutions. NaOH has been used as the reacting agent and added to the mixed solution of chlorides to maintain the pH value at 11. The chemical reaction of the mixed solution has been performed at 80 °C for 6 h and then cooled to room temperature. The co-precipitation has been collected and washed several times using double-distilled water with intermediate heating at 100°C for 20 minutes to remove the by-products, mainly NaCl. The co-precipitated powder then collected in a petri dish and heated at 180°C for 30minutes and the as-prepared powder free from bi-products has been ensured from the absence of white powder of NaCl. The green colored product has been made into fine powder by grinding and pellet form of the fine powdered material has been annealed at 800 °C for 4h in air for structural phase stabilization. The chemical composition of the Ga doped $\alpha$-$Cr_2O_3$ samples ($Cr_{1.85}Ga_{0.15}O_3$, $Cr_{1.45}Ga_{0.55}O_3$, and $Cr_{1.17}Ga_{0.83}O_3$) have been determined based on the results from energy dispersive X-ray analysis. The $\alpha$-$Cr_2O_3$ sample of grain size (~ 20 nm) in same range of Ga doped samples has been prepared by 60 h mechanical milling of bulk $\alpha$-$Cr_2O_3$ oxide and sample details are presented in earlier work [6]. We have used the mechanical milled $\alpha$-$Cr_2O_3$ sample (denoted $\alpha$-$Cr_2O_3$Mh60) as a reference for the discussion of structure, magnetic and electronic properties in Ga doped $\alpha$-$Cr_2O_3$ samples.

The X-ray diffraction (XRD) measurements on the pellet shaped samples after annealing have been carried out at room temperature in the $2\theta$ range $20^o$-$80^o$ (step size 0.02°) using Cu $K_\alpha$ radiation ($\lambda$=1.5406 Å) from X-ray diffractometer (Model: X-Pert PANalytical) with power generator set at 45 kV and 30 mA. Micro-Raman spectra of the samples have been recorded in the wave number range100 to 1000 $cm^{-1}$using Raman microscope (Renishaw, UK). The Raman spectra have been recorded at 514 nm radiation (2.41 eV) by exciting an Argon-Ion laser source optimized at 5 mW power. Dc magnetization has been recorded in the temperature range 5 K-350 K and magnetic field range ± 70 kOe using physical properties measurement system (PPMS-EC2, Quantum Design, USA). The zero field cooled (ZFC) and field cooled (FC) modes of magnetization measurements have been followed. In ZFC mode, the sample have been cooled from 350 K to low temperature (say, 10 K) in the absence of external magnetic field (H) or in the presence of suitable compensating magnetic field to maintain magnetization of the sample close to zero during cooling process. Then, magnetization measurement started in the presence of set magnetic field while temperature has been warmed up to 350 K. In FC mode, the sample has been cooled from 350 K under a constant magnetic field down to low set temperature (say, 10 K) and magnetization measurement started while the sample has been warmed up to 350 K without changing the cooling field. The magnetic field dependent magnetization [M(H)] curves have been measured at selected temperatures following ZFC mode. The M(H) loop in FC mode has also been recorded at 10 K after cooling the sample from 300 K in the presence of 70 kOe to examine the exchange bias effect. The Current-Voltage (I-V) characteristic curves of the samples have been measured using Keithley Source Meter (Model: 2410-C). The pellet shaped samples ($\varnothing \sim$ 13 mm, t ~ 1 mm) have been sandwiched between two platinum (Pt) electrodes using homemade sample holder (pt/sample/pt) and pressure technique. The I-V curves have been measured by sweeping the voltage from 0 to ± 20 V across the sample. Optical absorption spectra of the samples have been recorded in the UV-VIS range of light.

## 3. Results and discussion

### 3.1. Structural properties

Fig.1 has compared the XRD pattern of Ga doped $\alpha$-$Cr_2O_3$ samples ($\alpha$-$Cr_{2-x}Ga_xO_3$) with standard JCPDS- 0381479 XRD pattern of $\alpha$-$Cr_2O_3$ oxide. XRD pattern of the Ga doped samples have matched with rhombohedral structure (space group R$\bar{3}$C) of $\alpha$-$Cr_2O_3$. The sample with low Ga content ($Cr_{1.85}Ga_{0.15}O_3$) showed an unidentified extra peak at about 27°, where as the samples with relatively high Ga content ($Cr_{1.45}Ga_{0.55}O_3$ and $Cr_{1.17}Ga_{0.83}O_3$) formed single phase structure. The cell parameters (lattice constants: *a, c,* and cell volume (V)) of the samples have been determined from XRD profile fitting to rhombohedral structure. The extra peak for $Cr_{1.85}Ga_{0.15}O_3$ sample has been excluded during profile fitting. The cell parameters ($a$ = 4.964 Å, $c$ = 13.598 Å, V = 290.2 Å$^3$) of $Cr_{1.85}Ga_{0.15}O_3$ sample are found to be slightly larger in comparison to single phased $Cr_{1.45}Ga_{0.55}O_3$ ($a$ = 4.955 Å, $c$ = 13.572 Å, V = 288.5 Å$^3$) and $Cr_{1.17}Ga_{0.83}O_3$ ($a$ = 4.955 Å, $c$ = 13.573 Å, V = 288.7 Å$^3$) samples. The decrease of lattice parameters in the samples with higher Ga content is expected due to replacement of $Cr^{3+}$ ions with larger radius (0.65 Å) by $Ga^{3+}$ ions with smaller radius (0.62 Å) in rhombohedral structure.

The grain size ($<d>$) and lattice strain ($\varepsilon(\%)$) of the samples have been calculated using Williamson-Hall equation [8]: $\beta_{eff}cos\theta c = \frac{k\lambda}{<d>} + 2\varepsilon sin\theta$, where $\beta_{eff}^2 = \beta^2 - \beta_0^2$, $k$ is Scherrer constant (~0.9), $\varepsilon$ is the lattice strain, $\lambda$ is the x-ray wavelength. The peak position ($2\theta_c$) and full width at half maximum (FWHM: $\beta$) values have been calculated by fitting of 4-5 prominent XRD peaks using voigt shape. $\beta_0$ is FWHM for the peak of standard Si powder and used to correct the instrument broadening effect on XRD peaks. A linear extrapolation of the $\beta_{eff}cos\theta_c$ vs. $2sin\theta_c$ plot (Fig.1(d-f)) has been used to calculate $<d>$ (intercept on $\beta_{eff}cos\theta$ axis gives $k\lambda/<d>$) and $\varepsilon$ (slope). The grain size and lattice strain for the low Ga content $Cr_{1.85}Ga_{0.15}O_3$ sample ($<d>$~ 27 nm, $\varepsilon$ ~ 0.19 %) are found to be less than that in samples with higher Ga content ($Cr_{1.45}Ga_{0.55}O_3$: $<d>$ ~50 nm, $\varepsilon$ ~ 0.52 %and $Cr_{1.17}Ga_{0.83}O_3$: $<d>$ ~ 44 nm, $\varepsilon$ ~ 0.43 %). The smaller lattice strain in $Cr_{1.85}Ga_{0.15}O_3$ sample may be consistent to the fact that a fraction of the substituted Ga atoms wandering in the interstitial spaces, instead of occupying the lattice sites of Cr atoms. This creates defects/cation vacancies in the lattice structure of $Cr_{1.85}Ga_{0.15}O_3$ sample.

Raman spectroscopy has been used as a complementary technique for structural phase confirmation. The vibration modes in $\alpha$-$Cr_2O_3$ crystal (rhombohedral structure, $R\bar{3}C$ space group) are characterized by $2A_{1g}$, $2A_{1u}$, $3A_{2g}$, $2A_{2u}$, $5E_g$ and $4E_u$ bands. The $2A_{1g}$ and $5E_g$ bands are only Raman active [15, 20]. Fig.2 shows that Raman spectra of Ga doped $\alpha$-$Cr_2O_3$ samples are identical with a typical spectrum of $\alpha$-$Cr_2O_3$. The most intense peak positions of $A_{1g}$ band (~553 cm$^{-1}$) and $E_g$ bands (~306, 352, 397, 529, 615 cm$^{-1}$) in Ga doped $\alpha$-$Cr_2O_3$ samples are in agreement with Raman active modes of $\alpha$-$Cr_2O_3$ at about 307, 350, 395, 526, 551 and 610 cm$^{-1}$. This is the microscopic evidence of the incorporation of Ga atoms into the lattice sites of Cr atoms in $\alpha$-$Cr_2O_3$. The appearance of extra peaks (* marked) re-confirms structural defects only for the sample with low Ga (x =0.15) content. The profile parameters (position, width, height) of the major peaks have been obtained by fitting the peak profiles using Lorentzian shape. The peak position ($\bar{\nu}$) in Raman spectrum is associated with the variation of spring constant ($k_{MO}$) and effective mass ($\mu_{MO}$) of the M-O bonds by the relation $\bar{\nu} = \frac{1}{2\pi c}\sqrt{\frac{k_{MO}}{\mu_{MO}}}$. The $\mu_{GaO}$ is expected to be larger than $\mu_{CrO}$ due to larger atomic mass of Ga atom (69.723 a.u) than the atomic mass of Cr atom (51.996 a.u). On the other hand, spring constant ($k_{M-O}$), being second order derivative of the coulomb potential ($V(r_{MO})$), is proportion to $\frac{1}{r_{MO}^3}$ and depends on the ionic radii difference of $Ga^{3+}$ (0.62 Å) and $Cr^{3+}$ (0.65 Å) ions. Peak position of the bands (Fig. 2(e)) noticeably shifted with respect to Ga doping in $\alpha$-$Cr_2O_3$ oxide. The decrease of both lattice parameters and position of Raman active bands suggests that the effect of the change of effective mass is more prominent than the effect of the change of bong strength in Ga doped $\alpha$-$Cr_2O_3$ oxide. Fig. 2(f) shows that peak width of the low frequency bands ($E_g(1)$, $E_g(2)$), representing the displacement of the cations along perpendicular direction of rhombohdral plane, decreases with the increase of Ga content in $\alpha$-$Cr_2O_3$ structure, in contrast to the increasing peak width of the high frequency bands ($A_{1g}$, $E_g(5)$), generally associated with the displacement of cations along in-plane directions of the rhombohdral structure. The increase of band width of the peaks is generally associated with increasing local disorder in the lattice structure. In the present samples, in-plane lattice disorder most probably increases by the doping of Ga in $\alpha$-$Cr_2O_3$ structure. Fig. 2(g) shows a general tendency of increasing peak intensity, which is in contrast to the decrease of peak position ($\bar{\nu}$) with the increase of Ga content in $\alpha$-$Cr_2O_3$ structure, for most of the bands. The intensity of the scattered light (polarization direction $\mu$) in Raman spectrum is proportional to fourth power of peak position ($\bar{\nu}$) and square of the molecular polarizability $\alpha_{\mu\lambda}(\bar{\nu})$ for incident light with polarization direction $\lambda$ [11]. This means the change of molecular polarizability ($\alpha_{\mu\lambda}(\bar{\nu})$) due to displacement or re-adjustment of the Cr/Ga ions in Ga doped $\alpha$-$Cr_2O_3$ structure may dominate in determining the peak intensity in Raman spectra of Ga doped samples. The change of $\alpha_{\mu\lambda}(\bar{\nu})$ is important for magneto-optic and opto-electronic properties in $\alpha$-$Cr_2O_3$ structure [21].

## 3.2. Magnetic properties

Fig. 3(a-d) compared the temperature dependence of ZFC magnetization ($M_{ZFC}$) and FC magnetization ($M_{FC}$) curves measured at 100 Oe for mechanical milled $\alpha$-$Cr_2O_3$ sample (denoted by $Cr_2O_3$Mh60 with grain size $\sim$ 20 nm) and Ga doped $\alpha$-$Cr_2O_3$ samples. The $M_{ZFC}$ and $M_{FC}$ curves showed bifurcation below 300 K for all the samples. The observation of peak in $M_{ZFC}$ and $M_{FC}$ curves near to 288 K ($T_{m1}$) may be considered as the Neel temperature for $\alpha$-$Cr_2O_3$ (nanoparticle) sample. This peak temperature at $T_{m1}$ is located below the Neel temperature ($T_N \sim$ 320 K) of the bulk $\alpha$-$Cr_2O_3$ oxide. The results suggest the reduction AFM order in $\alpha$-$Cr_2O_3$ nanoparticle (our case, $Cr_2O_3$Mh60) sample [22, 23]. The temperature dependence of magnetic features of the samples can be explained by core-shell model for AFMNP, where core consists of spins with bulk AFM order and shell consists of frustrated (disordered) spins [2]. On lowering the temperature (reduction of thermal effect), $M_{ZFC}$ curve decreases due to AFM ordering of core spins and increase of $M_{ZFC}$ below 100 K indicates the superparamagnetic type contribution of frustrated/uncompensated shell spins of AFMNP [2]. In contrast, $M_{FC}$ curve of the $Cr_2O_3$Mh60 sample increased on lowering the temperature below $T_{m1} \sim$ 288 K. Interestingly, a kink (or minor peak) was observed at about 50 K ($T_{m2}$) for both $M_{ZFC}$ and $M_{FC}$ curves. Such low temperature kink in the temperature dependence of magnetization curve indicates the existence of second magnetic transition in $\alpha$-$Cr_2O_3$Mh60 sample. Such minor peak appears as an effect of the dilution of AFM spin order and a competitive effect between surface anisotropies [23, 24]. The decrease of magnetization below $T_{m2}$ can arise due to freezing or blocking of a fraction of interacting

surface (shell) spins, where as superparamagnetic type rise of magnetization at low temperature is exhibited by the majority of the frustrated (less interactive) surface spins. It may be noted that magnetic features of the Ga doped $\alpha$-$Cr_2O_3$ samples are remarkably different from $\alpha$-$Cr_2O_3$Mh60 sample. Both $M_{ZFC}$ and $M_{FC}$ curves in Ga doped $\alpha$-$Cr_2O_3$ samples monotonically increased on decreasing the temperature below 350 K without any visible peak near to 288 K, as noted for $\alpha$-$Cr_2O_3$Mh60 sample. These typical features, along with reduced gap between $M_{FC}$ and $M_{ZFC}$ curves, confirmed the dilution of AFM order by doping of non-magnetic Ga in $\alpha$-$Cr_2O_3$ structure. The AFM ordering temperature at $T_{m1}$ is either lost or small AFM background is over shadowed by the enhancement of ferromagnetic component in the system [7, 24]. Similar enhancement of ferromagnetic component in Ga doped $\alpha$-$Fe_2O_3$ (hematite) system has been observed below room temperature (AFM state of $\alpha$-$Fe_2O_3$) due to the increase of net uncompensated spin moment between neighboring (alternatively AFM spin ordered) rhombohedral planes [11]. In the present case, magnetization curves down to 100 K are relatively less temperature dependent for the samples with higher Ga content ($Cr_{1.45}Ga_{0.55}O_3$, $Cr_{1.17}Ga_{0.83}O_3$) in comparison to a continuous increase of magnetization below 250 K in the less Ga content $Cr_{1.85}Ga_{0.15}O_3$ sample. The $Cr_{1.85}Ga_{0.15}O_3$ sample showed defective lattice structure, where all the Ga atoms do not sit in the rhombohedral lattice structure of $\alpha$-$Cr_2O_3$. The temperature dependence of the inverse of ZFC magnetic susceptibility ($\chi = M_{ZFC}/H$) have been plotted in Fig. 3(e-h) to understand the modified magnetic order in Ga doped $\alpha$-$Cr_2O_3$ structure. The $\chi^{-1}(T)$ curve for the samples $\alpha$-$Cr_2O_3$ above 300 K and $Cr_{1.85}Ga_{0.15}O_3$ above 100 K are fitted with a straight line. This indicates the validity of Curie-Weiss law: $\chi = \frac{c}{T-\theta_p}$, applicable in paramagnetic state of a material. The large negative paramagnetic Curie temperature ($\theta_p \sim$ -102 K for $Cr_2O_3$Mh60 and -430 K for $Cr_{1.85}Ga_{0.15}O_3$) suggests AFM spin order still existing in the samples. However, $\chi^{-1}(T)$ curves (ZFC) for the samples with higher Ga content ($Cr_{1.45}Ga_{0.55}O_3$ and $Cr_{1.17}Ga_{0.83}O_3$) above 100 K are best fitted with an equation: $\frac{1}{\chi} = \frac{T-\theta_1}{C} - \frac{b}{T-\theta_2}$, applicable for ferrimagnet where magnetic spins form two unequal sub-structure [25]. The negative value of $\theta_1$($\sim$ 6060 K for $Cr_{1.45}Ga_{0.55}O_3$ and 7770 K for $Cr_{1.17}Ga_{0.83}O_3$) takes into account a strong AFM interaction in the samples. Although such a large negative value of $\theta_1$, obtained from the best fitting of the experimental data, may not be the exact due to involvement of four constant parameters in the fitting process, but the nature of $\chi^{-1}$(T) curves confirmed ferrimagnetic spin structure. The existence of second magnetic sub-structure with weak AFM interaction is taken care by the small negative value of $\theta_2$ ($\sim$ 23 K for $Cr_{1.45}Ga_{0.55}O_3$ and 115 K for $Cr_{1.17}Ga_{0.83}O_3$). There are two possibilities for the existence of two magnetic sub-structures either due to core-shell structure of nanoparticles [4, 25, 26] or due to unequal magnetic contribution arising from the differences of population of non-magnetic Ga and magnetic Cr ions in rhombohedral planes [9]. Similar ferrimagnetic properties have been reported for CrMnGaC sample [27]. The effective paramagnetic moment ($\mu_{eff}$) of Cr atoms has been estimated using the fit values of Curie constant ($C$) in the formula $\mu^2{}_{eff} = \frac{3k_B C}{N}$, where $k_B$ is the Boltzmann constant, N is the number of Cr atoms per unit gram of the sample. The calculated $\mu_{eff}$ per Cr atom ($\sim$ 0.41$\mu_B$, 2.50$\mu_B$ and 2.62$\mu_B$ for x = 0.15, 0.55 and 0.83, respectively in $Cr_{2-x}Ga_xO_3$) shows higher value for the single phased samples with higher content of Ga in α-$Cr_2O_3$ structure and the values are close to 2.61 $\mu_B$ calculated for $Cr_2O_3$Mh60 sample.

Now, we proceed to understand the low temperature magnetic features of the Ga doped samples. Fig. 4(a-c) shows the $M_{ZFC}$(T) curves in at different magnetic fields. We marked the low temperature freezing/blocking transition at $T_{m2}$ for $Cr_{1.45}Ga_{0.55}O_3$ and $Cr_{1.17}Ga_{0.83}O_3$ samples. The first order derivative of $M_{ZFC}$(T) curves, normalized by room temperature data, in Fig. 4(d-f) confirms the existence of low temperature spin freezing temperature in all Ga doped samples. In addition to an increase of magnetization with magnetic field, spin freezing temperature shows an unusual variation with magnetic field (inset of Fig. 4(d-f)). The freezing temperature ($T_{m2}$) initially increased with magnetic field (typically in the range 1 kOe-10 kOe). At higher magnetic field, $T_{m2}$ of the samples showed a usual decrease with the increase of magnetic field. It seems that two spin relaxation processes control magnetic dynamics in Ga doped samples depending on the competition between magnetic anisotropy and interactions in the AFM system [4, 28-29]. In the first process, a fraction of spin attempts to restore the original AFM order of the core through magnetic exchange coupling at the interfaces of core and shell. Such spin dynamics in AFM system dominates at low magnetic fields [28-29], showing the increase of freezing temperature ($T_{m2}$). After reaching a critical interfacial interaction and effective volume of AFM spin cluster, the magnetic field is used to overcome the anisotropy barriers of the interacting spins clusters. This results in a usual decrease of the freezing temperature of the spin clusters on further increase of magnetic field. In order to understand the interactions at the interfaces of core and shell spins at low temperature and at different fields, the $M_{ZFC}(T)$ curves below 50 K have been analyzed by a modified Curie type equation $\chi_{sp}(T) = \frac{M_{ZFC}}{H} = \frac{C_{sp}}{T^{\alpha}}$, generally used for AFMNP system [2], where $\chi_{sp}$ and $C_{sp}$ are the susceptibility and Curie constant, respectively contributed by frustrated surface spins. The exponent ($\alpha$) estimates the interactions among surface spins and at the interfaces of core and shell. Fig. 5(a) shows the fitted data according to modified Curie type equation of the samples at different fields. Fig. 5(b) shows that $\alpha$ varied in the range 0.08-0.13, 0.40-0.32 and 0.59-0.39 for the samples $Cr_{1.85}Ga_{0.15}O_3$, $Cr_{1.45}Ga_{0.55}O_3$ and $Cr_{1.17}Ga_{0.83}O_3$, respectively for measurement magnetic field in the range 100 Oe-70 kOe. The $\alpha$ value is expected to be 1 for non-interacting spins in ideal superparamagnetic or paramagnetic state and less than 1 for a system having finite interactions among the surface spins [2, 10]. The sample $Cr_{1.85}Ga_{0.15}O_3$ exhibited relatively low $\alpha$ values and variation of $\alpha$ with magnetic field (decreasing trend only at higher fields) is characteristically different from the samples with higher Ga content (decreasing trend throughout the fields). The magnitude and nature of the variation of $\alpha$ with magnetic field indicate that magnetic interactions among the surface spins at lower temperatures is stronger for the low Ga content (defective) sample in comparison to the samples with higher Ga content.

The magnetic field dependence of magnetization (M(H)) curves in Fig. 6 (a-c) re-confirms a strong magnetic interaction at 10 K for the (defective) $Cr_{1.85}Ga_{0.15}O_3$ sample (broad M(H) loop with higher coercivity $\sim$ 1715 Oe) in comparison to single-phased samples (small loop with low coercivity $\sim$ 245 Oe for $Cr_{1.45}Ga_{0.55}O_3$ and $\sim$ 45 Oe for $Cr_{1.17}Ga_{0.83}O_3$ samples). The FC-M(H) loop measured at 10 K after field cooling from 300 K under 70 kOe field showed a larger shift towards negative field axis in case of $Cr_{1.85}Ga_{0.15}O_3$ sample than that in $Cr_{1.45}Ga_{0.55}O_3$ sample. The exchange bias field, i.e., shift of the center of FC loop with respect to the center of ZFC loop, was found $\sim$90 Oe and 50 Oe for $Cr_{1.85}Ga_{0.15}O_3$ and $Cr_{1.45}Ga_{0.55}O_3$, respectively. Such exchange bias shift in AFMNPs suggests the magnetic exchange interactions between the core-shell spins [2, 26, 30-31]. The first order derivative of the magnetization curves (dM/dH) at 10 K (Fig. 6(d-f)) exhibited a butterfly-wing shaped hysteresis with peaks near to the coercivity of the samples. Such typical feature indicates the coexistence of FM and AFM orders below the spin freezing temperature ($T_{m2}$) of the samples [24]. The butterfly-shaped loop of dM/dH vs. H curve at 10 K becomes field independent at 300 K (inset of Fig. 6(d-f)). This confirms paramagnetic state of the Ga doped samples at 300 K, where M(H) curve showed linear behaviour. The magnetic field dependence of dM/dH (dynamic susceptibility) curve also suggests a gradual change in the magnetic spin order at 10 K, showing the existence of a stronger FM component in $Cr_{1.85}Ga_{0.15}O_3$ (defective) sample and a strong AFM component in Ga doped $\alpha$-$Cr_2O_3$ structure with higher Ga content, where magnetic field induced spin order showed a non-linear variation of dynamic susceptibility without butterfly-shaped loop. We demonstrate the basic features of the semiconductor nature of the material using current-voltage curve and optical absorption of the structurally single phased samples ($\alpha$-$Cr_2O_3Mh60$, $\alpha$-$Cr_{1.45}Ga_{0.55}O_3$, $\alpha$-$Cr_{1.17}Ga_{0.83}O_3$).

### 3.3 Current-voltage (I-V) characteristics

The current-voltage (I-V) characteristics curve was measured by sandwiching the samples between two Pt electrodes (Pt/Sample/Pt) and sweeping the electric voltage within ± 20 V in four segments with positive bias sequence (PBS) $0 \rightarrow +20\,V \rightarrow 0 \rightarrow -20\,V \rightarrow 0 \rightarrow +20\,V$ (inset of Fig. 7(a)) and negative bias sequence (NBS) $0 \rightarrow -20\,V \rightarrow 0 \rightarrow +20\,V \rightarrow 0 \rightarrow -20\,V$ (inset of Fig. 7(d)). The measurement of I-V curves in 4 segments was repeated for 20 times. The I-V curves of the samples are shown in Fig. 7 (a-c), Fig. 7 (d-f) and Fig. 7 (g-i) for the PBS, NBS and repetition for 20 times, respectively. In the PBS (Fig. 7 (a-c)), $\alpha$-$Cr_2O_3Mh60$ sample showed a small I-V loop for higher voltages of positive and negative biases. The I-V loop is comparatively wide, starting from the lower voltage in both positive and negative biases, for the sample $\alpha$-$Cr_{1.45}Ga_{0.55}O_3$. On the other hand, I-V loop is also comparatively wide, starting from the lower voltage only in PBS, for the sample $\alpha$-$Cr_{1.45}Ga_{0.55}O_3$. For this sample, the loop in the NBS is extremely narrow, even smaller than that in $\alpha$-$Cr_2O_3Mh60$ sample. The basic features of the I-V loop of the samples using NPS (Fig.7(d-f)) are well consistent to that observed in PBS mode, except some specific changes. For example, relatively smaller loop observed during paths 1 and 2 of positive voltage than that during paths 3 and 4 (negative voltage) in PBS mode has reversed by the relatively wider loop during paths 1 and 2 (positive voltage) than that during paths 3 and 4 (negative voltage) in NBS mode for the sample $\alpha$-$Cr_{1.45}Ga_{0.55}O_3$. The I-V loop behavior confirms the existence of two irreversible electronic states (bi-stable states: low resistance state (LRS) and high resistance state (HRS)) in the samples. Recently, similar electronic hysteresis loop has been reported for Ga doped hematite ($\alpha$-$Fe_2O_3$) system [12] and many other systems [32-39]. In case of $\alpha$-$Cr_2O_3$Mh60 sample, the system is electronically in HRS during increase of positive voltage (PBS-1) that switched to LRS during the decrease of positive voltage (return path PBS-2). On changing the polarity of bias voltage from positive to negative, the resistance state switched from LRS during increase of negative voltage (path PBS-3) to HRS during the decrease of negative voltage (return path PBS-4). Such electric field induced switching of resistance state (HRS to LRS and LRS to HRS) on changing the polarity of bias voltage in electronic devices is called bipolar resistance switching [34]. On the other hand, Ga doped samples showed abnormal bipolar switching property, where resistance state switched from LRS (during PBS-1) to HRS (during PBS-2) and retained the same switching property (i.e., LRS to HRS) on changing the polarity of bias voltage. The switching property of the samples as noted in PBS is also preserved in NBS mode of measurement. Fig. 7(g-i) confirms the reproducibility of I-V loop features of the samples by repeating the measurement for 20 times using PBS, except the magnitude of current slightly decreases on increasing the number of repetition. The decrease of current at initial state (within 4-5 repetitions) is quite rapid due to relaxation of the charge carriers (see in the inset of Fig. 7(g-i)). The current values are well stabilized for higher number of repetitions. Interestingly, $Cr_{1.45}Ga_{0.55}O_3$ sample exhibited NDR (negative differential resistance) effect for bias voltage above 5 V and such unusual resistance switching behavior is not observed for other two samples. Technologically, bi-stable electronic states in magnetic semiconductors are applicable in the field of memory and switching devices [33, 35], including logic circuit, frequency multiplier and divider, voltage-controlled oscillator, and flip-flop circuit [30].

Although an exact mechanism of the bi-stable electronic states is not understood in highly resistive materials, the non-linear electronic properties in many electronic devices, where an insulator/semiconductor sample is sandwiched between two metal electrodes, are significantly affected by the extrinsic charge carriers [32-33]. The general belief is that a competition between conduction mechanisms, i.e., electrode limited conduction and bulk limited conduction, produce irreversible electronic states (LRS and HRS) and NDR effect [ 36-39]. The intrinsic ionic defects and vacancy controlled trapping and de-trapping of the charge carriers at the interfaces constitute electrode limited conduction, whereas transport of the charge carriers through the insulator/semiconductor sample constitutes the bulk conduction. We analyzed the non-linear I-V curves of the samples in the forward and return paths of the positive bias mode (PBS-1 and PBS-2) using the equations that described the charge transfer mechanism at the interfaces of metal electrode-semiconductor-metal electrode frame work. The space charge field at the interfaces reduces the current flow through the sample (bilk limited conduction). The modified Child's law $I \sim V^m$ can be applied to represent the space charge limited current (SCLC) mechanism. The exponent ($m = \frac{\Delta lnI}{\Delta lnV}$ the slope of the lnI vs. lnV plot) is expected nearly 2 for SCLC mechanism in solid state material [33-34]. We observed that power law behavior may be applicable for the lower limit of applied voltage of $Cr_2O_3Mh60$ sample with the $m$ values ~ 1.14 and 1.4 for paths PBS-1 and PBS-2, respectively (fitting shown in the insets of Fig. 8(a)). The $m$ values ~ 0.94 and 1.11 for $Cr_{1.45}Ga_{0.55}O_3$ sample (Fig. 8(c)), and ~ 2.21 and 2.02 for $Cr_{1.17}Ga_{0.83}O_3$ sample (Fig. 8(e)) were found at low voltage limits of the paths PBS-1 and PBS-2, respectively. Even at higher voltage limit, the $m$ values were found in the range 1.46-1.62, well below of 2. Hence, I-V curves in $Cr_2O_3Mh60$ and $Cr_{1.45}Ga_{0.55}O_3$ samples may not be controlled by SCLC mechanism, but SCLC mechanism may dominate the I-V characteristics for $Cr_{1.17}Ga_{0.83}O_3$ sample. On the other hand, electric field controlled conduction mechanism at the interfaces of electrode and semiconductor, where electric field overcomes the energy barriers at the interfacial junctions, can be understood from Schottky emission formula $J = AT^2 \exp[\frac{-q(\varphi_B - \sqrt{\frac{qV}{4\pi\varepsilon d}})}{k_B T}]$. The Poole-Frenkel emission formula $J = qN_c\mu E \exp[\frac{-q\varphi_B}{k_B T} + \frac{q\sqrt{\frac{qV}{\pi\varepsilon d}}}{rk_B T}]$ is the bulk analogue of Schottky effect at the interfacial barrier [33]. Here, J is the current density (I per unit cross sectional area), A is the Richardson constant, T is absolute value of measurement temperature, $q\varphi_B$ is the barrier height or trap energy level at the interface, V voltage, $k_B$ Boltzmann's constant, $\varepsilon$ is the dynamic dielectric constant (usually less than static value of dielectric constant), d is the sample thickness, q is the electronic charge, $N_C$ is the density of charge carriers in the conduction band, and $\mu$ is the mobility of charge carriers. We used the return path of the positive bias mode (PBS-2) to analyze the I-V curve using emission equations (Fig. 8 (b, d, g)). The Poole-Frenkel mechanism and Schottky mechanism can be verified by the linear fit of the data in ln(I/V) vs. $\sqrt{V}$ plot and lnI vs. $\sqrt{V}$ plot, respectively with slope 's' values shown in the insets of Fig. 8(b, d, g). We have observed that Schottky mechanism and Poole-Frenkel mechanism are applicable in the higher and lower limit of voltages, respectively. A detailed analysis of the I-V characteristics will be published elsewhere.

### 3.4. Optical absorption properties

We recorded the optical absorption spectra of the samples in UV-visible range of light to confirm electronic transitions from the occupied state of valence band (low energy state) to the unoccupied state of conduction band (high energy state) by absorbing the incident electro-magnetic radiation, where tailing of the band edges due to defects will be reflected in the nature of the optical spectra. The transitions between two bands can be either direct (crystal momentum is conserved and electronic transition is sharp) or indirect (phonon mediated transition and electronic transition occurs over a range of wavelengths of incident radiation). Thus, a minimum energy is required for making such electronic transitions depending on the band structure. Band gap energy of the material can be estimated by identifying the threshold wavelength in the optical absorption spectra. Fig. 9(a) shows absorption spectra of the samples in UV-visible range of wavelength of the incident radiation. The absorption spectra clearly showed threshold wavelengths at around 410 nm and 535 nm, respectively. The threshold wavelength for excitation of electrons from valance band to conduction band is not well defined for wavelengths above 600 nm. On the other hand, three broad absorption peaks in the visible-UV spectrum at wavelengths around 365 nm (peak 1), 460 nm (peak 2) and 600 nm (peak 3), respectively indicated three prominent electronic transitions in $\alpha$-Cr$_2$O$_3$Mh60 and Ga doped $\alpha$-Cr$_2$O$_3$ samples. The observed peaks are consistent to earlier reports of $\alpha$-Cr$_2$O$_3$ [22, 26]. The peak at about 365 nm is associated with the conventional band gap ($\sim$ 3 eV) of $\alpha$-Cr$_2$O$_3$ [13]. The absorption peaks at around 460 nm and 600 nm correspond to $^4$A $_{2g}$ $\rightarrow$ $^4$T$_{1g}$ and $^4$A$_{2g}$ $\rightarrow$ $^4$T$_{2g}$ 3d$^3$ electronic transitions of octahedral Cr$^{3+}$ ions [22]. It is difficult to classify whether the peaks are due to direct or indirect electronic transitions for wide band gap semiconductors, like $\alpha$-Cr$_2$O$_3$. In such case, Tauc relation: $\alpha h\nu = A(h\nu - E_g)^n$ [13] has been used to calculate the band gap (E$_g$). The exponent $n$ = 1/2 and 2 are applicable for direct and indirect electronic transitions, respectively. We noted $n$ = 1/2 gives a better linear extrapolation in the $(\alpha h\nu)^2$ vs. h$\nu$ plot (Fig. 7(b)) for calculating the band gap at $\alpha$ = 0 and obtained values of band gap are shown in the inset of Fig. 7(b). The optical band gap in $\alpha$-Cr$_2$O$_3$ structure has been decreased by Ga doping. The Cr$_{1.85}$Ga$_{0.15}$O$_3$ sample showed slightly less band gap, most probably due to defective structure [14], in comparison to single phased samples with high Ga content.

## 4. Conclusions

The present work demonstrates the modifications in the lattice structure, magnetic and electronic properties as an effect of Ga doping in $\alpha$-$Cr_2O_3$ system. The lattice structure has stabilized in rhombohedral structure with space group $R\overline{3}C$. The observation of an extra phase indicated defective lattice structure for low Ga content $Cr_{1.85}Ga_{0.15}O_3$ sample. The defective structure has shown characteristically different magnetic properties (large magnetic loop and exchange bias shift at 10 K) and optical absorption properties (smaller band gap) in $Cr_{1.85}Ga_{0.15}O_3$ sample in comparison with the single phased samples with higher Ga content. The doping of non-magnetic Ga atoms in $\alpha$-$Cr_2O_3$ system has diluted AFM spin order. The enhancement of ferromagnetic component has exhibited the signature of ferrimagnetic properties in the temperature dependence of magnetization curves for the samples with higher Ga content. The low temperature magnetic features (spin freezing and superparamagnetism) appeared due to contribution of the interfacial spins of core-shell structure of the AFMNP system (grain size range 27-50 nm). However, proper incorporation of the non-magnetic Ga atoms in $\alpha$-$Cr_2O_3$ structure does not show major changes of magnetic moment per Cr atom in the system. The non-linear I-V curves of the samples showed two irreversible electronic states LRS and HRS along with loop behaviour. The space charge limited current mechanism may be affecting the nature of the I-V curves up to certain extent, but the prominent mechanism for the exhibition of hysteresis loop and NDR effect could be related to the charge accumulation and transfer across the interfaces of metal electrode-semiconductor junctions (electrode limited conduction) and transport of charge carriers through the material (bulk limited conduction). Over all, the modified magnetic, electronic and optical absorption properties as the effect of non-magnetic Ga doping in AFM semiconductor oxide $\alpha$-$Cr_2O_3$ can find potential applications in the field of spintronics.

**Acknowledgments**

Authors thank to CIF, Pondicherry University for recording Raman spectra and EDX data. RNB acknowledges research grants from Department of Science and Technology, Ministry of Science and Technology (No. SR/S2/CMP-0025/2011) for carrying out this work.

**Figure captions**

Fig .1 XRD pattern of $\alpha$-$Cr_2O_3$ and Ga doped $Cr_2O_3$ samples (a-d) and linear fit of the XRD peak parameters ($\beta_{eff}$ and $\theta_C$) using Williamson-Hall equation in (e-g).

Fig. 2 Raman spectra of the Ga doped $Cr_2O_3$ samples (a-d). The peaks profiles are fitted using Lorentzian shape. The obtain values of peak position, FWHM and peak height of the samples for selected bands are plotted with Ga concentration (e-g).

Fig. 3 Temperature dependence of magnetization (a-d) and inverse of the ZFC dc susceptibility (e-h) for $\alpha$-$Cr_2O_3$ and Ga doped $Cr_2O_3$ samples, measured at 100 Oe field.

Fig. 4 Temperature dependence of magnetization (a-c) and first order derivative of the normalized magnetization curves (d-f) at different magnetic fields for the Ga doped samples. The inset shows the spin freezing temperature at different fields.

Fig. 5 Fit of the $H/M_{ZFC}$ vs. $T^{\alpha}$ data at low temperatures (a) and fitted values of $\alpha$ at different magnetic fields for the Ga doped samples.

Fig. 6 Magnetic field dependence of magnetization at 10 and 300 K (a-c), ZFC and FC loops at 10 K (inset of a-b), dM/dH vs H plots at 10 K and 300 K (d-f) for Ga doped $Cr_2O_3$ samples.

Fig. 7 I-V curves in positive bias sequence (a-c), in negative bias sequence (d-f) and 20 times repetition (g-i) for the samples $Cr_2O_3Mh60$, $Cr_{1.45}Ga_{0.55}O_3$ and $Cr_{1.17}Ga_{0.83}O_3$, respectively. The bias sequences within $\pm$ 20 V are shown in the inset of (a) and (d). The variations of current during 20 times repetition for the samples are shown in the insets of (g), (h) and (i).

Fig.8 Analysis of the experimental data in I-V curves PBS-1 and PBS-2 for samples $Cr_2O_3Mh60$ (a-b), $Cr_{1.45}Ga_{0.55}O_3$ (c-d) and $Cr_{1.17}Ga_{0.83}O_3$ (e-f) according to different mechanisms, as shown in the insets.

Fig. 9 UV-vis absorption spectra (a) and corresponding Tauc plot (b) for $\alpha$-$Cr_2O_3$ and Ga doped $Cr_2O_3$ samples. Three prominent absorption peaks and associated band gaps of the sample are indicated by arrows.

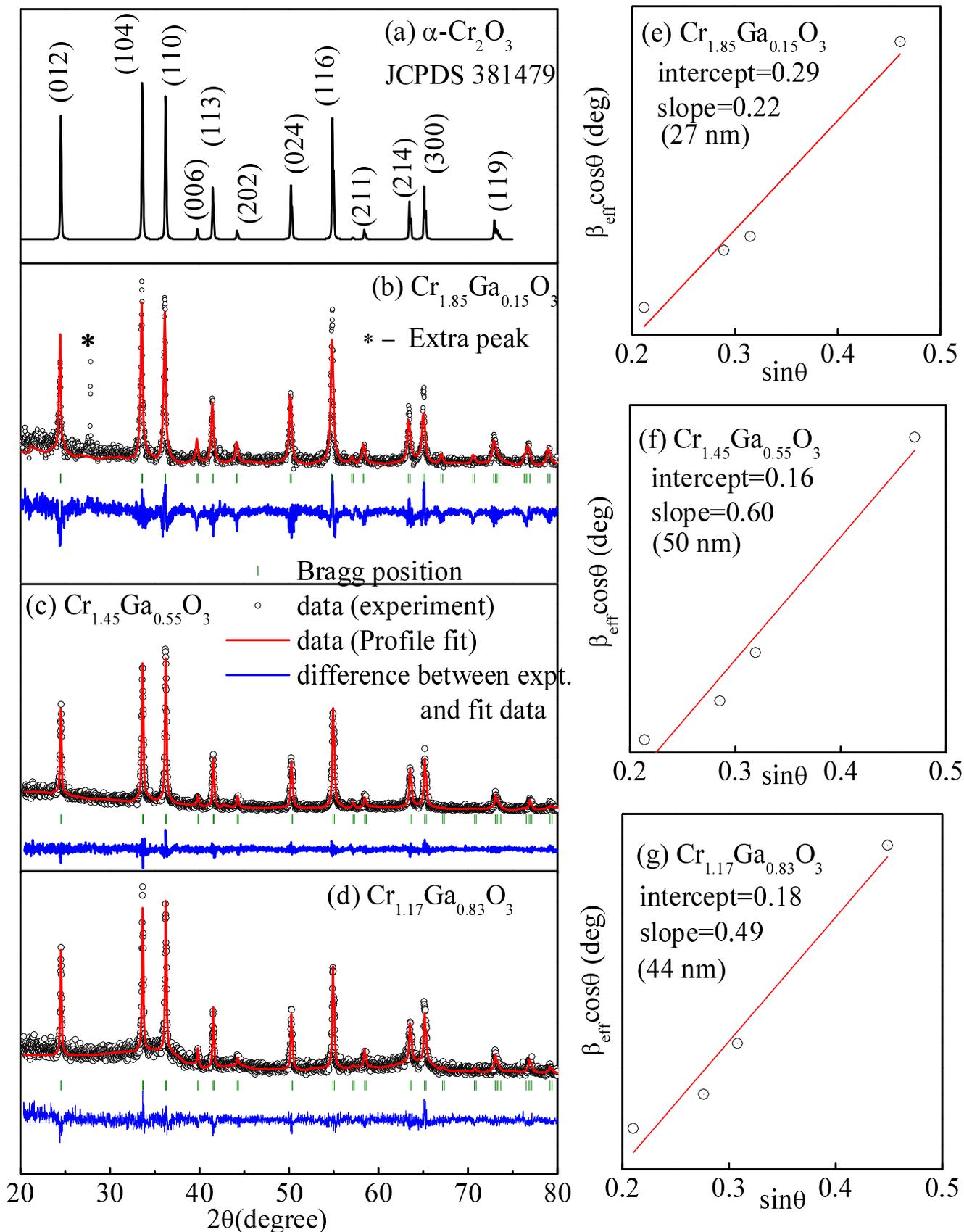

Fig . 1 XRD pattern of $\alpha$-$Cr_2O_3$ and Ga doped $Cr_2O_3$ samples (a-c) and linear fit of the XRD peak parameters ($\beta_{eff}$ and $\theta_C$) using Williamson-Hall equation in (d-f).

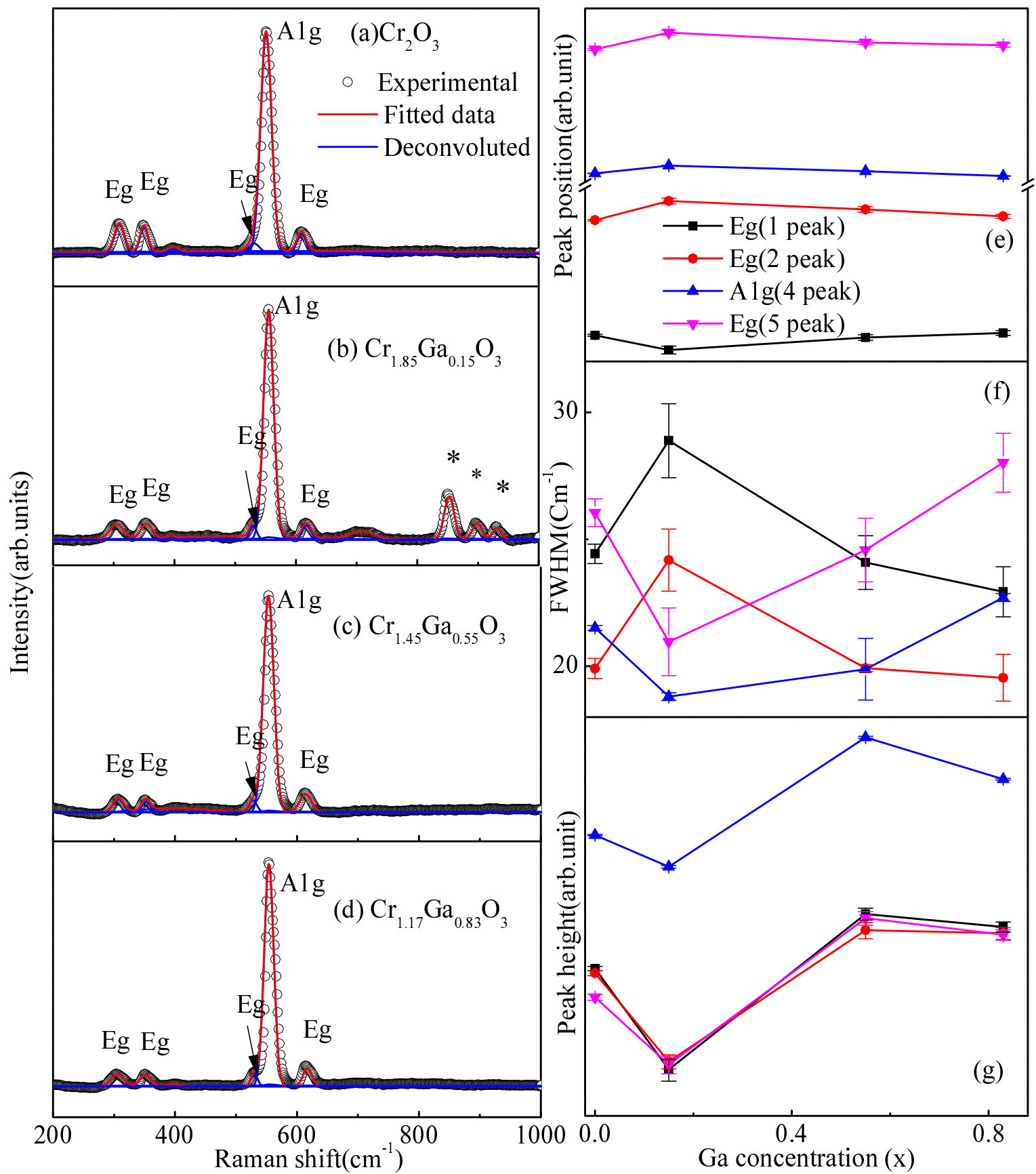

Fig. 2 Raman spectra of the Ga doped $Cr_2O_3$ samples (a-d). The peaks profiles are fitted using Lorentzian shape. The obtain values of peak position, FWHM and peak height of the samples for selected bands are plotted with Ga concentration (e-g).

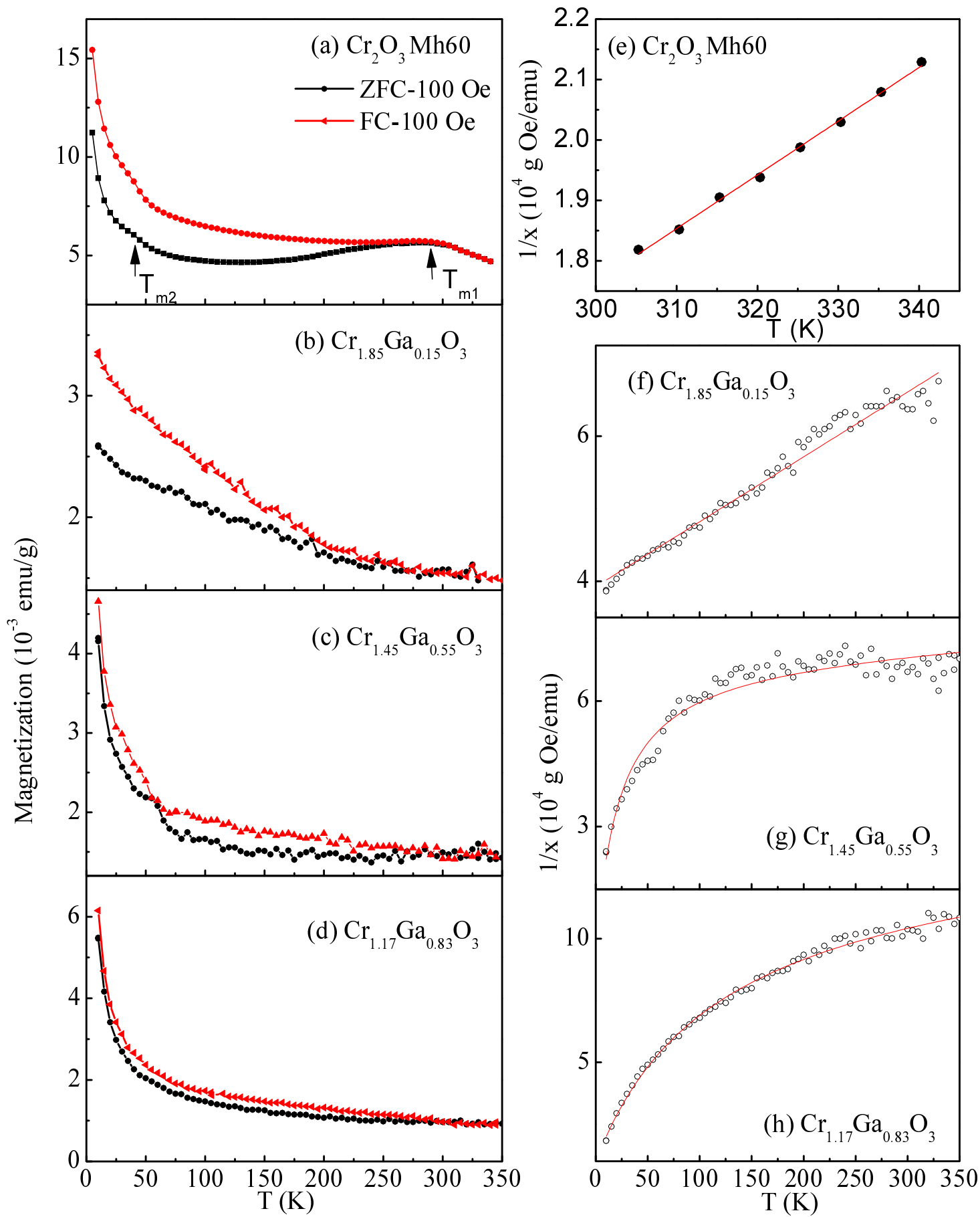

Fig. 3 Temperature dependence of magnetization (a-d) and inverse of the ZFC dc susceptibility (e-h) for $Cr_2O_3$ and Ga doped $Cr_2O_3$ samples, measured at 100 Oe field.

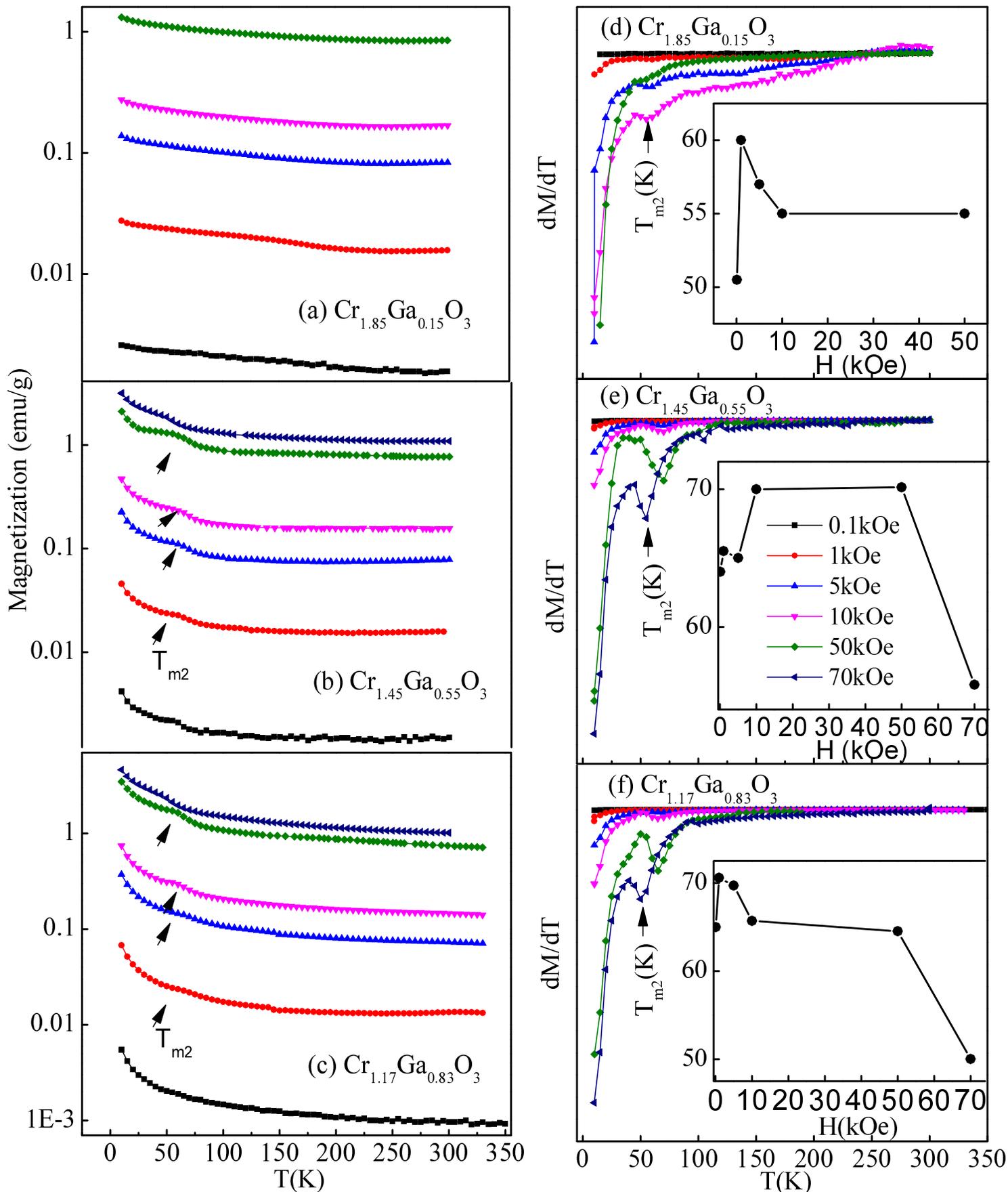

Fig. 4 Temperature dependence of magnetization (a-c) and first order derivative of the normalized magnetization curves (d-f) at different magnetic fields for the Ga doped samples. The inset shows the spin freezing temperature at different fields.

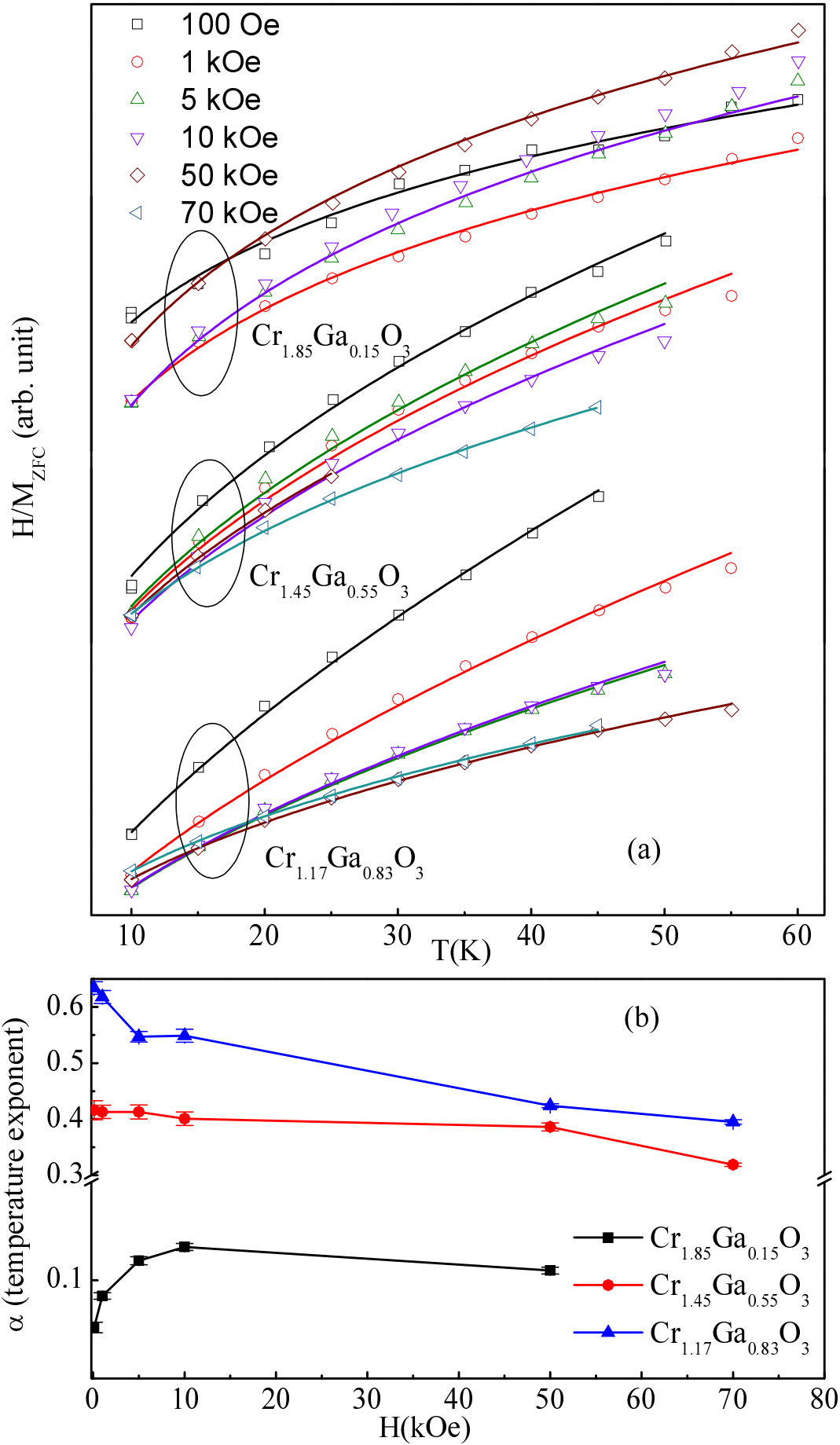

Fig. 5 Fit of the H/M$_{ZFC}$ vs T$^\alpha$ data at low temperatures (a) and fitted values of $\alpha$ at different magnetic fields for the Ga doped samples.

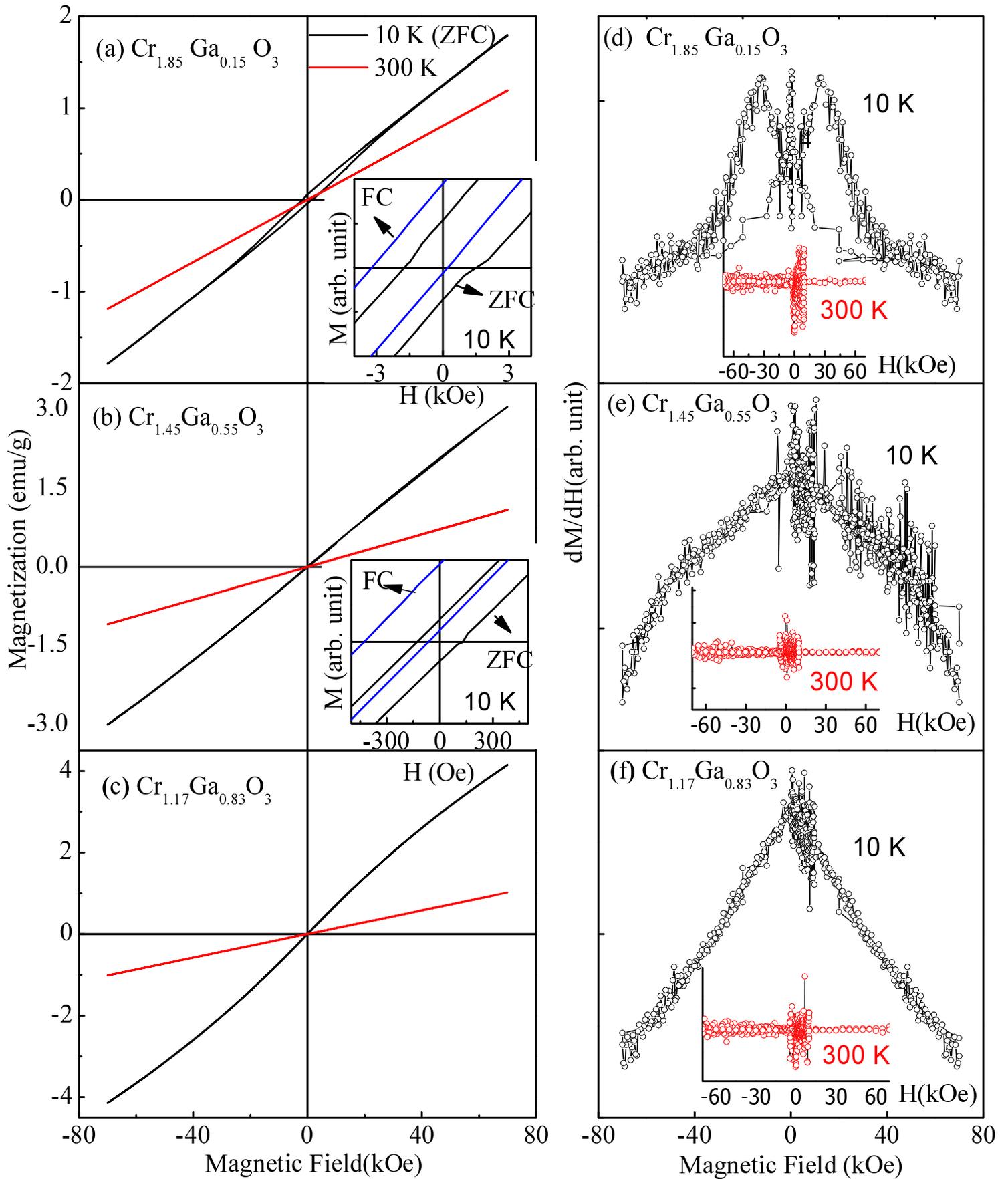

Fig. 6. Magnetic field dependence of magnetization at 10 and 300 K (a-c), ZFC and FC loops at 10 K (inset of a-b), dM/dH vs H plots at 10 K and 300 K (d-f) for Ga doped $Cr_2O_3$ samples.

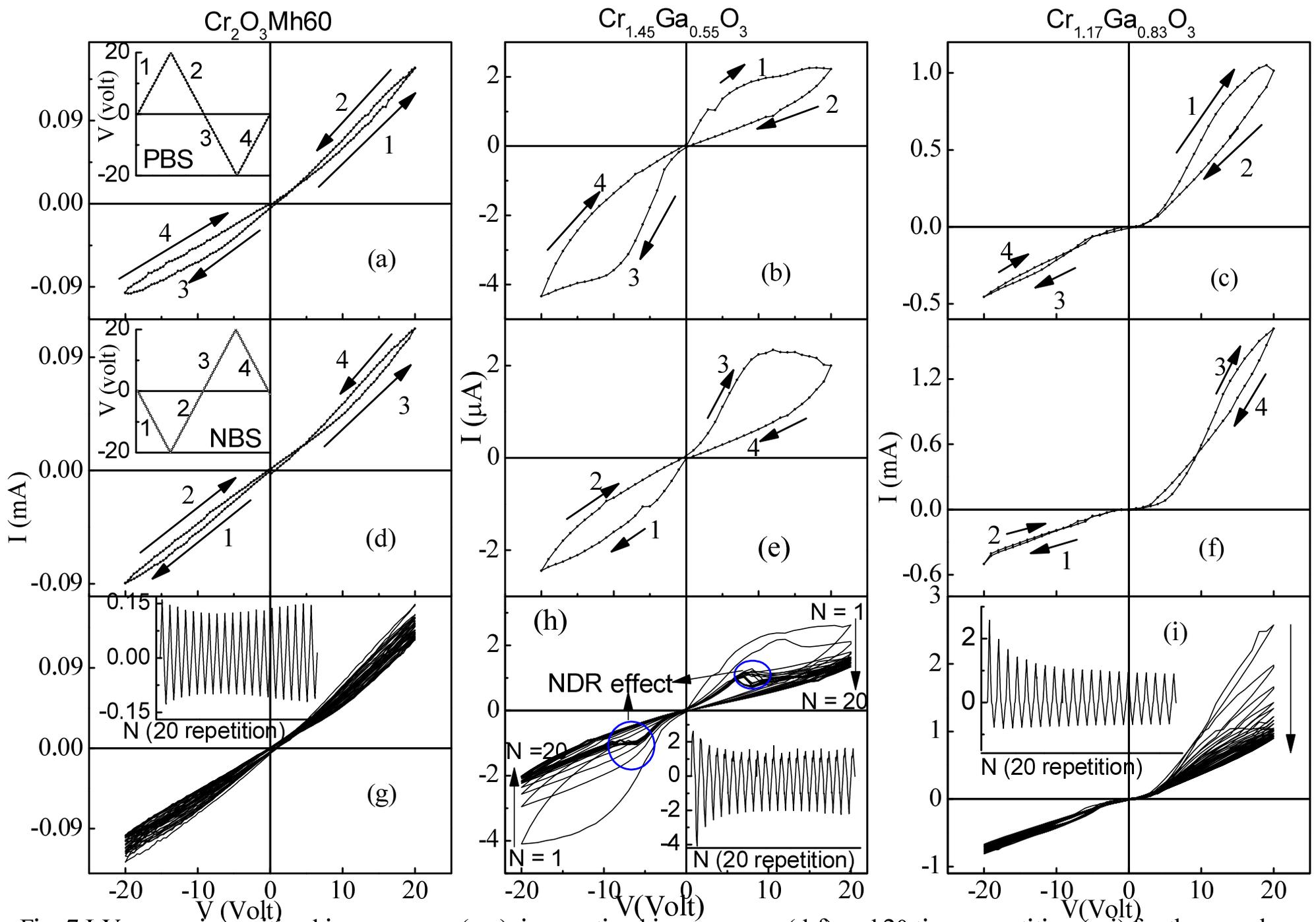

Fig. 7 I-V curves in positive bias sequence (a-c), in negative bias sequence (d-f) and 20 times repetition (g-i) for the samples $Cr_2O_3Mh60$, $Cr_{1.45}Ga_{0.55}O_3$ and $Cr_{1.17}Ga_{0.83}O_3$, respectively. The bias sequences within $\pm 20$ V are shown in the inset of (a) and (d). The variations of current during 20 times repetition for the samples are shown in the insets of (g), (h) and (i).

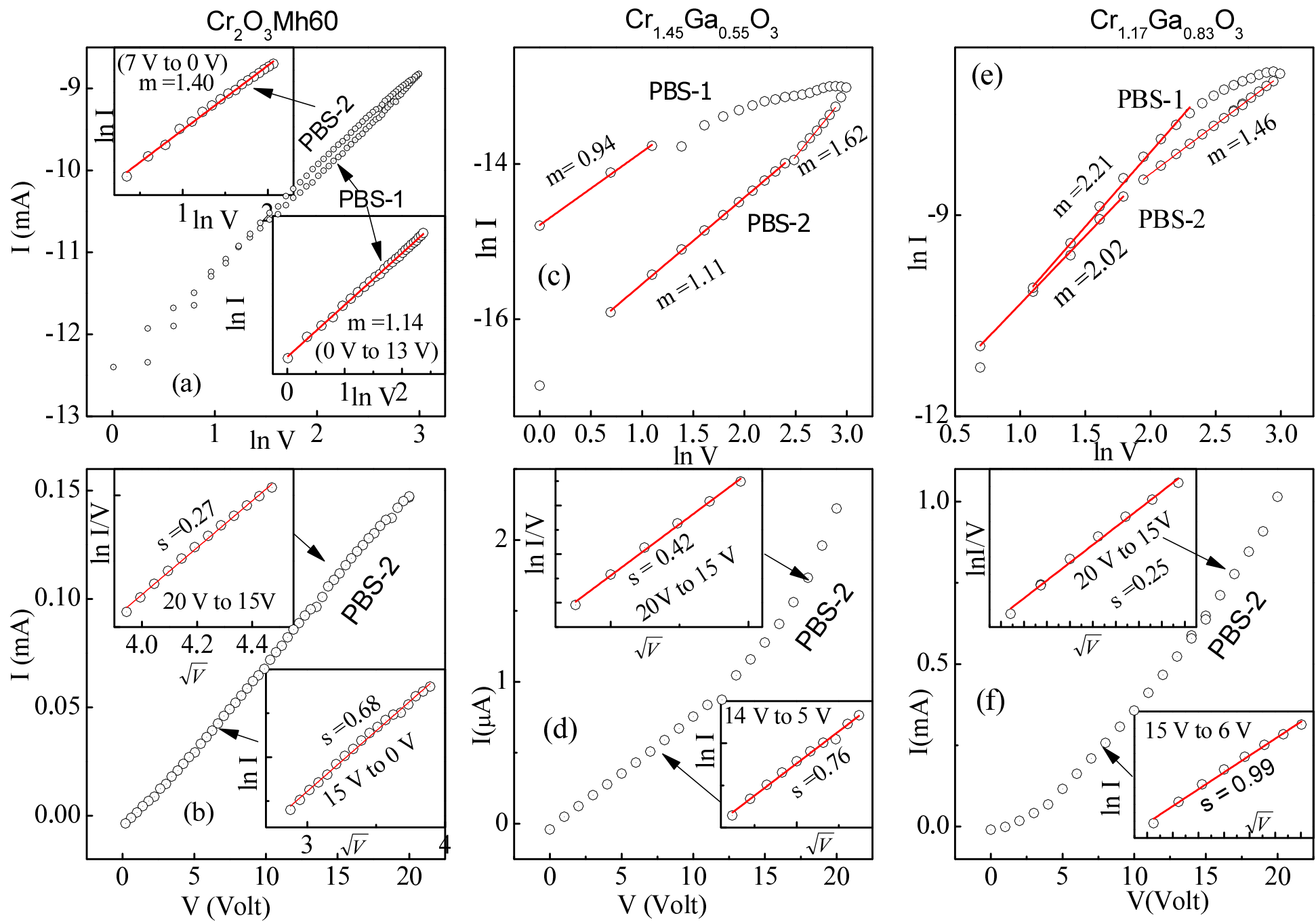

Fig.8 Analysis of the experimental data in I-V curves PBS-1 and PBS-2 for samples $Cr_2O_3Mh60$ (a-b), $Cr_{1.45}Ga_{0.55}O_3$ (c-d) and $Cr_{1.17}Ga_{0.83}O_3$ (e-f) according to different mechanisms, as shown in the insets.

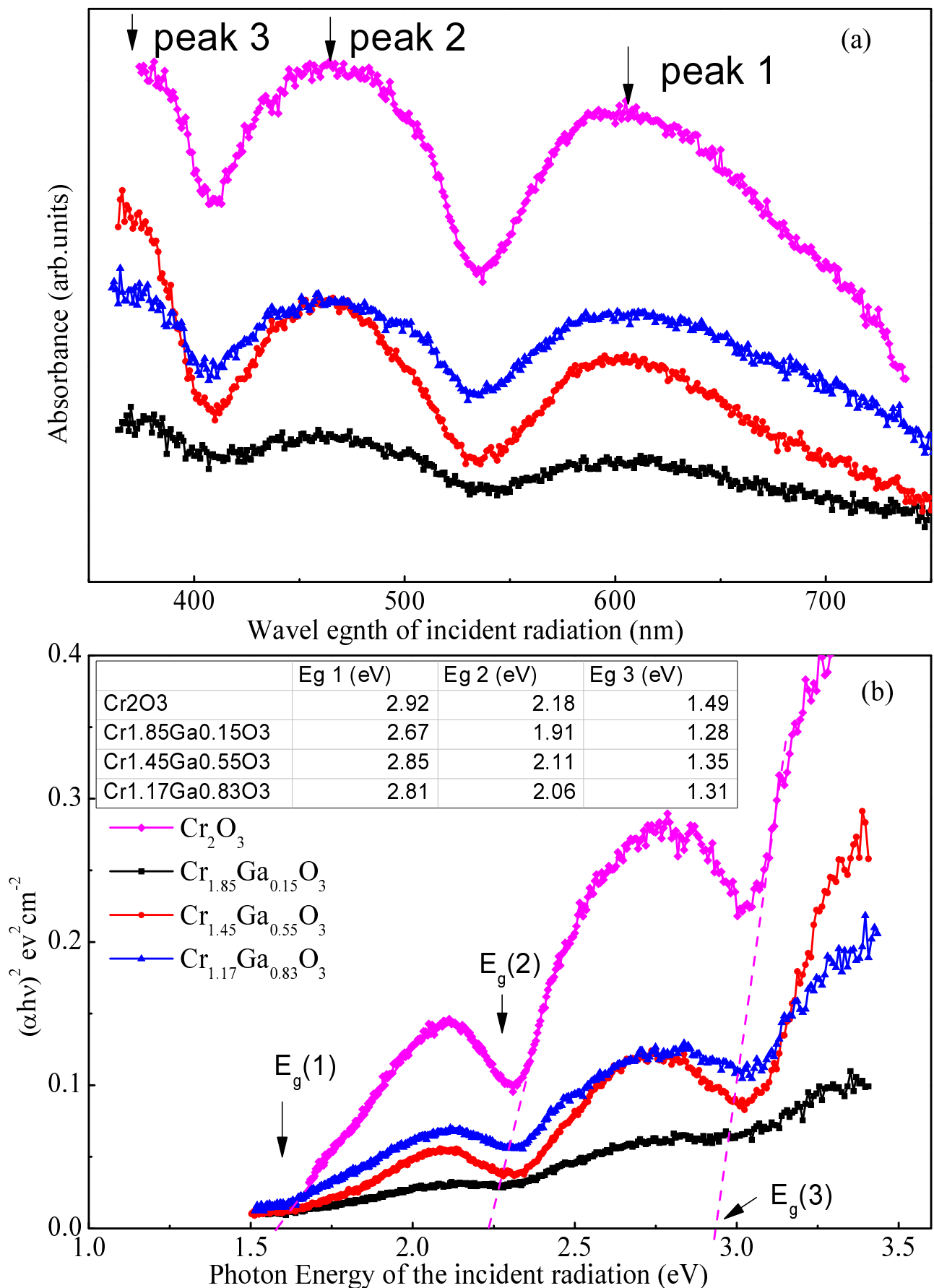

Fig. 9 UV-vis absorption spectra (a) and corresponding Tauc plot (b) for α-Cr$_2$O$_3$ and Ga doped Cr$_2$O$_3$ samples. Three prominent absorption peaks and associated band gaps of the sample are indicated by arrows.